\documentclass[conference]{IEEEtran}
\IEEEoverridecommandlockouts
\usepackage[hidelinks]{hyperref}
\usepackage{flushend}
\usepackage{cite}
\usepackage{amsmath,amssymb,amsfonts}
\usepackage{algorithmic}
\usepackage{graphicx}
\usepackage{textcomp}
\usepackage{xcolor}
\usepackage{booktabs} \def\BibTeX{{\rm B\kern-.05em{\sc i\kern-.025em b}\kern-.08em
    T\kern-.1667em\lower.7ex\hbox{E}\kern-.125emX}}

\usepackage{listings}
\usepackage{xcolor}

\definecolor{codeblue}{rgb}{0.22, 0.45, 0.70}
\definecolor{codegreen}{rgb}{0.0, 0.5, 0.0}
\definecolor{codered}{rgb}{0.6, 0, 0}
\definecolor{codegray}{rgb}{0.5, 0.5, 0.5}

\lstdefinestyle{java}{
  language=Java,
  basicstyle=\ttfamily\footnotesize,
  keywordstyle=\color{codeblue}\bfseries,
  stringstyle=\color{codered},
commentstyle=\color{codegreen}\itshape,
  numbers=left,
  numberstyle=\tiny\color{codegray},
  stepnumber=1,
  numbersep=10pt,
  backgroundcolor=\color{white},
  showspaces=false,
  showstringspaces=false,
  tabsize=4,
  breaklines=true,
  breakatwhitespace=true,
  frame=single,
  rulecolor=\color{codegray},
  captionpos=b,
  escapeinside={(*@}{@*)},
      abovecaptionskip=10pt
}

\lstdefinestyle{console}{
basicstyle=\ttfamily\footnotesize, 
    backgroundcolor=\color{gray!20}, 
    frame=single,
        frame=none, 
    breaklines=true,
    keepspaces=true,
    columns=fixed,
    deletekeywords={new, for},
    abovecaptionskip=10pt,
        numbers=none 
}

\usepackage{float}
\usepackage{placeins}
\usepackage{fancybox,framed}
\usepackage[utf8]{inputenc}

\usepackage[figure,norelsize,vlined,linesnumbered]{algorithm2e}
\SetAlFnt{\small}
\SetAlCapFnt{\large}
\SetAlCapNameFnt{\large}

\SetCommentSty{mycommfont}

\usepackage{booktabs} 
\usepackage{xspace}
\usepackage{tabularx,colortbl}
\usepackage{multirow, makecell}
\usepackage{paralist}
\usepackage{tabularx}
\usepackage{listings}
\usepackage{graphicx}
\usepackage{ifthen}
\usepackage{amsthm}
\usepackage{indentfirst}
\usepackage{xcolor}

\usepackage{pifont}
\definecolor{darkjunglegreen}{rgb}{0.000000,0.392157,0.000000}

\usepackage{url}
\def\codesize{\small}
\def\<#1>{\codeid{#1}}\protected\def\codeid#1{\ifmmode{\mbox{\codesize\ttfamily
			#1}}\else{\codesize\ttfamily
		#1}\fi}

\definecolor{mygray}{gray}{0.9}
\usepackage{fancybox}
\usepackage{listings}
\usepackage{etoolbox}
\usepackage{kvoptions}
\makeatletter
\newcommand{\removelatexerror}{\let\@latex@error\@gobble}
 
\usepackage{xspace}
\newcommand{\tool}{\textsc{LLMloop}\xspace}
\newcommand{\humaneval}{\textsc{HumanEval-X}\xspace}

\newcommand{\java }{\textsc{Java}\xspace}

\newcommand{\evosuite }{\textsc{EvoSuite}\xspace}

\SetKwProg{Fn}{Function}{}{}
\makeatother

\usepackage{booktabs} 
\usepackage{graphics} 
\usepackage{multirow,bigstrut}
\usepackage[framemethod=tikz]{mdframed}
\usepackage{graphicx}
\usepackage{bm} 
\usepackage{relsize}

\usepackage{bold-extra}

\DeclareMathAlphabet\mathbfcal{OMS}{cmsy}{b}{n}

\usepackage{eso-pic}
\usepackage{multirow,array,varwidth}
\usepackage{booktabs}
\usepackage{pifont}

\usepackage{listings}
\usepackage{color,url}

\definecolor{javared}{rgb}{0.6,0,0} 
\definecolor{javagreen}{rgb}{0.25,0.5,0.35} 
\definecolor{javapurple}{rgb}{0.5,0,0.35} 
\definecolor{javadocblue}{rgb}{0.25,0.35,0.75} 
\usepackage{lipsum}
\usepackage{paralist}

\newtheoremstyle{mystyle}
{0.0mm}
{0.0mm}
{}
{}
{\bfseries}
{.}
{ }
{}

\usepackage{amsbsy}
\definecolor{orange}{rgb}{1,0.5,0}
\definecolor{darkjunglegreen}{rgb}{0.000000,0.392157,0.000000}

\usepackage{soul}

\usepackage{multicol}

\font\tt=rm-lmtl10

\usepackage{enumitem}
\usepackage{amsmath}
\usepackage{microtype}
\usepackage{amsthm}

\newlength\marincrease



\hypersetup{
  pdftitle={LLMLOOP: Improving LLM-Generated Code and Tests Through Automated Iterative Feedback Loops},
  pdfauthor={Ravin Ravi; Dylan Bradshaw; Stefano Ruberto; Gunel Jahangirova; Valerio Terragni},
  pdfsubject={This paper introduces LLMLOOP, an automated iterative feedback-loop approach for refining LLM-generated code and tests using tool-based feedback to improve quality and correctness.},
  pdfkeywords={AI4SE; software testing; program synthesis;
automated test generation; Large Language Models}
}

\begin{document}

\title{\tool: Improving LLM-Generated Code and Tests through Automated Iterative Feedback Loops}
\author{
\centering
\IEEEauthorblockN{
Ravin Ravi\IEEEauthorrefmark{1}, 
Dylan Bradshaw\IEEEauthorrefmark{1}, 
Stefano Ruberto\IEEEauthorrefmark{2}, 
Gunel Jahangirova\IEEEauthorrefmark{3}, and
Valerio Terragni\IEEEauthorrefmark{1}
}
\IEEEauthorblockA{\IEEEauthorrefmark{1}University of Auckland, Auckland, New Zealand\\
\{rrav468, dbra157, vter674\}@aucklanduni.ac.nz}
\IEEEauthorblockA{\IEEEauthorrefmark{2}JRC European Commission, Ispra, Italy\\
 Stefano.RUBERTO@ec.europa.eu}
\IEEEauthorblockA{\IEEEauthorrefmark{3}King's College London, London, United Kingdom\\
gunel.jahangirova@kcl.ac.uk}
}

\maketitle
\AddToShipoutPictureFG*{%
  \AtPageLowerLeft{%
    \raisebox{1.6cm}{%
      \hspace*{3.0cm}%
      \parbox{\dimexpr\paperwidth-5.0cm\relax}{
      This is the authors’ version of the paper published in IEEE International Conference on
      Software Maintenance and Evolution (ICSME 2025).
      DOI: \href{https://doi.org/10.1109/ICSME64153.2025.00109}{10.1109/ICSME64153.2025.00109}.}%
    }%
  }%
}

	\begin{abstract}
Large Language Models (LLMs) are showing remarkable performance in generating source code, yet the generated code often has issues like compilation errors or incorrect code. Researchers and developers often face wasted effort in implementing checks and refining LLM-generated code, frequently duplicating their efforts.

This paper presents \tool, a framework that automates the refinement of both source code and test cases produced by LLMs. \tool employs five iterative loops: resolving compilation errors, addressing static analysis issues, fixing test case failures, and improving test quality through mutation analysis. These loops ensure the generation of high-quality test cases that serve as both a validation mechanism and a regression test suite for the generated code.

We evaluated \tool on \humaneval, a recent benchmark of programming tasks. Results demonstrate the tool effectiveness in refining LLM-generated outputs. A demonstration video of the tool is available at \url{https://youtu.be/2CLG9x1fsNI}.
	\end{abstract}


\maketitle

\section{Introduction}


Recent years have seen remarkable progress in large language models (LLMs), leading to their adoption across a wide range of domains~\cite{hadi2024large}. Beyond generating textual responses, LLMs have shown great potential in generating source code~\cite{chen2021evaluating, liu2024your, ugare2024improving, du2024evaluating}, promising to improve the productivity of software engineers~\cite{terragni2025future}.

\smallskip
Researchers and practitioners are studying and evaluating the effectiveness of LLMs in generating source code from textual prompts. While the results are promising~\cite{liu2024your,hou2024systematic,chen2021evaluating},  several challenges remain~\cite{liu2024your,du2024evaluating}. For instance, such code often fails to compile, as LLMs do not inherently perform compilation checks, requiring more sophisticated pipelines to handle such tasks. Additionally, LLM-generated code frequently suffers from dependency issues~\cite{du2024evaluating}, such as missing libraries. These models are also limited by the quality and of their training data~\cite{perez2021automatic}, which may include outdated or buggy code samples. 

\begin{figure*}[th!]
    \centering
    \includegraphics[width=0.95\linewidth]{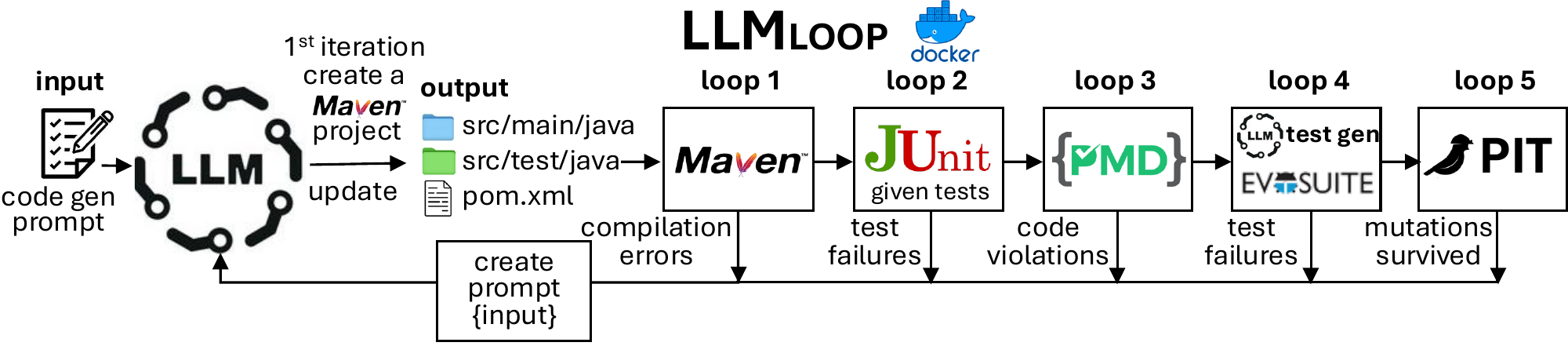}
\vspace{-3mm}
    \caption{Logical architecture of \tool}
    \label{fig:overview}
\end{figure*}

\smallskip
We observed that researchers and practitioners consistently encounter these issues. This leads to significant wasted effort in implement automated techniques to fix LLM-generated code.  
A commonly adopted strategy involves \textbf{feedback loops} that integrate LLMs with source code analysis tools (e.g., compilers)~\cite{quoc2024empirical}. These tools identify issues, which are then fed back to the LLM, providing contextual clues to refine the code based on the reported problems.
This iterative approach allows LLMs to refine their output, correct mistakes, and produce more robust and contextually appropriate code. However, setting up these feedback loops often requires significant effort to implement and setup properly.

\smallskip
This paper presents \textbf{\tool}, a framework to enhance the quality and reliability of LLM-generated \textsc{Java} code. \tool serves as an intermediary between developers/researchers and LLMs, providing automated mechanisms to fix, validate, and improve both source code and test cases produced by LLMs. The tool is fully automated and highly configurable, enabling researchers and developers to efficiently refine LLM-generated code. \tool avoids the repeated effort in implementing feedback loops and can be seamlessly integrated in any tool or workflow that uses LLMs to generate \textsc{Java} source code. 

\smallskip
The framework is for \textsc{Java} and runs within a dedicated \textsc{Docker} image, functioning as a secure sandbox environment. This ensures that the framework operates in isolation, protecting the host system from the execution of any potentially harmful code generated by the LLM during test case execution.

\smallskip
\tool currently implements five iterative self-refinement loops: (i) to fix compilation errors, (ii) to improve code quality issues reported by static analysis,  (iii) to fix failures of automatically generated test cases, (iv) to improve the quality of the test cases via mutation analysis. These test cases serve a dual purpose: identifying bugs in LLM-generated code and providing an associated test suite for the generated code.

\smallskip
We evaluated \tool by assessing its effectiveness in generating programming solutions of the \humaneval~\cite{chen2021evaluating} benchmark, as compared to a baseline approach that invokes the LLM only once (without incorporating any feedback loop). The results show that \tool improves the quality of LLM-generated code: pass@10 of 90.24\% versus pass@10 of 76.22\% for the baseline. We believe that \tool provides substantial benefits to the developer and research communities that are using LLMs to generate code. A demonstration video on how to use it is available at \texttt{\url{https://youtu.be/2CLG9x1fsNI}}

\smallskip
We publicly released \tool's code and experimental data: \texttt{\url{https://github.com/ravinravi03/LLMLOOP}}




\section{\tool}

Figure~\ref{fig:overview} illustrates the logical architecture of \tool, which leverages iterative feedback loops and multiple analysis methods to enhance LLM-generated \textsc{Java} code. For each type of feedback, \tool generates a dedicated prompt to guide the LLM in improving the code (production or test) based on the feedback (see ``create prompt \{input\}'' in Figure~\ref{fig:overview}).

\medskip
\textbf{Inputs.}
The inputs to \tool are: 1) a prompt to generate a program, 2) a series of command-line arguments (see Table~\ref{table:cli}), 3) an optional test suite to validate the generated code. 

\smallskip
The framework dynamically adjusts the LLM's temperature to optimize the generation process. It begins with the default deterministic setting (temperature = 0) or the temperature specified in input (flag \texttt{-t}) but increments the temperature by 0.1 in successive runs if errors persist. This introduces variability, potentially resolving repeated issues. Studies like Liu et al.~\cite{liu2024your} suggest an optimal temperature for generating code is around 0.2, guiding \tool's adaptive adjustments.


\medskip
\textbf{Outputs.}
The output of \tool is a \textsc{Maven} project containing the improved code generated by the LLM along with an automatically generated test suite in addition to the given test suite (if any). 

\medskip
While the canonical input is a prompt and arguments, \tool can also start from an existing \textsc{Maven} project. In such cases, it can automatically fix and improve a specific class under test or integrate newly generated code into the project. This functionality is particularly useful as it relies on dependencies already declared in the project. Additionally, users can use \tool to create a new project from scratch, fix bugs in an existing project, or enhance test suite coverage by providing the root directory and the path to a test class. For simplicity, the rest of the paper will assume the canonical use of \tool.

\medskip
\textbf{Implementation Details.}
\tool is implemented for \java programs, leveraging the \textbf{\textsc{Maven} ecosystem} and the \textsc{Maven} build automation tool for dependency resolution.

\smallskip
To address security concerns related to running LLM-generated code, \tool operates within a \textbf{\textsc{Docker}} container. This isolated environment safeguards the host system from potential vulnerabilities, including malicious code that might originate from open-source training data. Using a sandbox environment ensures the protection of sensitive information and maintains system integrity~\cite{wang2023review, chen2021evaluating}.


\smallskip
\tool's implementation uses the OpenAI APIs\footnote{\url{https://platform.openai.com/docs/models}} to interact with LLMs, leveraging OpenAI's extensive range of powerful LLMs and its position as a leading provider in the LLM field. However, it could be easily adapted to any LLM. 


\subsection{Feedback Loops in \tool}

\textbf{Loop 1: Fixing Compilation Errors}
The first feedback loop ensures all generated or modified code is compilable. The process begins with an initial prompt to the LLM, supplemented with comments to direct it to produce \textsc{Java} 11-compliant code. 
We choose \textsc{Java} 11 because \evosuite currently supports up to Java~11\footnote{\url{https://github.com/EvoSuite/evosuite}}. The LLM's response is structured as a nested JSON object containing:
\begin{itemize}
    \item \texttt{src}: A map of file paths to code strings.
    \item \texttt{main}: Configuration for running the project.
    \item \texttt{dependencies}: A list of required dependencies.
\end{itemize}

\medskip
The framework parses the JSON object, updates the \textsc{Maven} project, and attempts to compile it. If compilation errors are detected, their details (location and type) are fed back to the LLM for refinement. This loop continues until the code compiles successfully or a set limit is reached (\texttt{-n}).

\begin{table}[t]
\setlength{\tabcolsep}{3pt}
\renewcommand{\arraystretch}{1}
\centering
\caption{Command-line flags of \tool}
\rowcolors{1}{}{gray!10}
    
\resizebox{\linewidth}{!}{%
\begin{tabular}{cll|}
\toprule
\textbf{Flag} & \textbf{Additional Argument} & \textbf{Description} \\ \midrule
-e    & -                                 & Enable \evosuite  \\ 
-p    & Path to \textsc{Java} project (String)      & Path to an existing \textsc{Java} project  \\ 
-d    & -                                 & Enable debug logging  \\ 
-t    & -                                 & Enable test generation using LLMs  \\ 
-r    & -                                 & Enable code coverage report  \\ 
-n    & Number of retries (Int)           & Set the number of attempts in the feedback loop  \\ 
-s    & -                                 & Enable static analysis of code generated by LLM  \\ 
-mut  & -                                 & Enable mutation analysis of tests generated by LLM  \\ 
-temp & Temperature value (Float)         & Set the temperature of the LLM model  \\ 
-ts   & Path to test suite (String)       & The relative path of the test suite  \\ 
-depth & Dependency depth (Int)           & The depth of the dependency tree  \\ 
-m    & Number of minutes (Int)           & Number of minutes that \evosuite runs for  \\ \bottomrule
\end{tabular}
}

\label{table:cli}
\end{table}

\medskip
\textbf{Loop 2: Test Failures} After generating compilable code, the framework runs the given test cases (if any). The feedback provided to the LLM consists of the test cases that fail, along with the corresponding stack traces for each failure. In this loop, we assume the given tests are correct and only ask the LLM to fix the generated code so that the tests pass.

\medskip
\textbf{Loop 3: Static Analysis}
When all given test pass, or the budget is reached, the framework performs static analysis using the PMD \textsc{Maven} plugin.
PMD\footnote{https://docs.pmd-code.org/latest/index.html} is an extensible multilanguage static code analysis tools that detects common programming flaws like unused variables, empty catch blocks, unnecessary object creation, etc. It comes with 400+ built-in rules and can be extended with custom rules. It parses source files into abstract syntax trees (AST) and runs rules against them to find violations. Once the PMD report is generated, \tool performs the following steps:
\begin{enumerate}
    \item Parses the PMD report for violation details.
    \item Prompts the LLM to fix these issues.
    \item Iterates until no violations remain or the iteration limit is reached.
\end{enumerate}

\medskip
To ensure compatibility, the framework dynamically verifies and adds required plugins and dependencies to the \texttt{pom.xml} file using \textsc{Maven}.

\medskip
\textbf{Loop 4: Test Generation}
\tool supports two test generation strategies:
\begin{itemize}
    \item \textbf{\evosuite:} A search-based tool that generates unit tests with high code coverage~\cite{fraser2011evosuite,jahangirova2023sbfttrack}.
    \item \textbf{LLM-Generated Tests:} The LLM creates test cases based on provided prompts and project context~\cite{schafer2023empirical, liu2024llm, yu2023llm,kang2023large}.
\end{itemize}

\medskip
The two modes of test generation offer different approaches to testing. The test suite generation by \evosuite focuses on achieving a set of coverage objectives guided by a fitness function. The produced test cases contain test oracles that capture the implemented behaviour (but not the intended behaviour), leading to test cases applicable only for a regression scenario~\cite{konstantinou2024llms}, so all tests would pass by construction. 

\smallskip
The test generation by LLMs, however, is not guided by any pre-determined algorithm, but is expected to produce test code similar to the one produced by humans (due to being trained on a large set of human-produced data) and often produces test oracles that capture the expected behaviour instead of the implemented behavior~\cite{konstantinou2024llms,Ruberto2025FromImplemented}. The prompt used by \tool explicitly instructs to generate test cases that achieve high coverage, to ensure that both positive and negative scenarios are covered and to take into account exceptional cases and boundary values. 

\smallskip
Both types of generated test cases are validated through the \textsc{Maven} test lifecycle. If failures occur for the LLM-generated tests, the details are sent back to the LLM for refinement. This iterative process continues until the test suite pass all the test. Note that in order to make the test suites pass the LLMs can change either the generated source code, the generated test code, or both. 
This means that \tool lets the LLM to autonomously determine whether the test cases reveal a bug in the generated code or if the test cases themselves contain issues that need to be fixed. As such, loop 4 can potentially improve both the code under test and the associated automatically LLM-generated test suite.

\medskip
\textbf{Loop 5: Mutation Analysis}
Once \tool ensures that all the generated test cases are passing, it proceeds with the further improvement of the test suite quality.
For this purpose, we use mutation testing, which evaluates the test suite effectiveness by introducing errors (mutants) into the code.  These mutants are slight variations of the original code. The test suite is then executed against these mutants to determine whether it can detect the seeded errors. If a test suite fails to identify a mutant, it indicates a potential weakness or gap in the tests. 
This approach provides a quantitative measure of test suite quality. Loop 5 needs that all generated tests are passing on the given version, as mutation testing requires a `green` test suite were all test pass. Note that this step is only performed for the generated tests, not for the given tests (used in loop 2). 

 \tool uses the PIT\footnote{https://pitest.org/} mutation testing tool and performs the following steps:
\begin{enumerate}
    \item Generate a report of surviving and killed mutants.
    \item Provide details of surviving mutants to the LLM for refinement.
    \item Iterate until all mutants are addressed or the iteration limit is reached.
\end{enumerate}

\smallskip

\section{Evaluation}

\textbf{Benchmark.}  
We evaluated \tool with \textbf{\humaneval}~\cite{zheng2023codegeex}, an extension of the popular \textsc{HumanEval}~\cite{chen2021evaluating} benchmark. Unlike the original benchmark, which focuses solely on \textsc{Python}, \humaneval expands to multiple programming languages, including \java, and introduces 80$\times$ more test cases~\cite{liu2024your}. \humaneval contains \textbf{164 coding problems in \java}, each comprising a function signature, a docstring describing the function's intended behavior (which we will use as the initial LLM prompt), and a test suite of unit tests. These tests  serve as a widely accepted proxy for correctness. Following prior work on program synthesis, we consider a solution correct if it passes all the provided test cases~\cite{zheng2023codegeex}. The test suite is divided into example and validation tests. \textit{Example tests} (used in Loop~2) are designed to guide the program synthesis tool in solving the task; \textit{validation tests} are exclusively used to assess the correctness of the generated code and kept hidden during program synthesis.  

\medskip  
\textbf{Experimental Setup.}  
We developed a \textsc{Python} script to convert the function signatures and docstrings from \humaneval into class skeletons. Additionally, we transformed \humaneval's tests, originally written as standalone main functions, into \textsc{JUnit} tests for compatibility with the framework. After reformatting the problems, \tool launches a \textsc{Docker} container based on the framework's image, with the project files mounted into the container for execution.  

\smallskip
For our experiments, we used OpenAI's \textsc{GPT-4o-mini} LLM. The terminal prompt provided to the \textsc{Docker} container included all the necessary flags to perform every stage of the framework (see Table~\ref{table:cli}). The number of retries (\texttt{-n}) and the depth of the dependency tree (\texttt{-d}) were both set to five. We presented each of the 164 \textsc{Java} problems to the LLM using a standardized initial prompt for code generation. To account for the non-deterministic nature of LLM output, we ran \tool ten times for each problem.  

\smallskip
The second column in Table~\ref{tab:result} gives the order of loops used in our experiments. We started with the Compilation Loop (Loop 1 in Figure~\ref{fig:overview}) because successful compilation is necessary for executing tests and performing static analysis. We then used the \humaneval example tests to guide the feedback loop (Loop 2), applied static analysis (Loop 3), and finally generated additional tests using both LLM and \evosuite (Loops 4-5). It is important to mention that Loop 1 is triggered every time the code changes, even in subsequent loops. This is because Loop~1 must always ensure that the code is compilable to perform the static analysis and run the test suites.

\medskip  
We evaluated the framework’s effectiveness with \textbf{pass@k}~\cite{chen2021evaluating}: The probability that at least one out of \(k\) independently generated code samples for a given problem passes all the test cases. We computed \textit{pass@1} to \textit{pass@10}, aligned with the ten runs for each problem in input. 
 This metric is particularly relevant for LLMs due to their non-deterministic nature, where identical inputs may produce different outputs. Considering multiple samples, \textit{pass@k} provides a robust measure of performance~\cite{chen2021evaluating}.  

\begin{table}[t]
\setlength{\tabcolsep}{2pt}
\renewcommand{\arraystretch}{1.2}
\centering
\caption{Pass@1 of \humaneval problems (mean over 10 runs) at each step of \tool.}
\rowcolors{1}{}{gray!10}
    
\resizebox{\linewidth}{!}{%

\begin{tabular}{rlcc}
\hiderowcolors
\toprule
\centering
 & \textbf{\tool stage}               & \textbf{pass@1} & \textbf{pass@1} \\ 
              & \textbf{(mean $\pm$ std)} &  \textbf{(mean \%)} \\ \midrule 
              \showrowcolors
\#1 & Baseline  (no feedback loop)                   & 117.50 $\pm$ 1.20                                   & 71.65\%                  \\ 
\#2 & Compilation (Loop 1)             & 125.30 $\pm$ 1.55                                   & 76.40\%                  \\ 
\#3 & \humaneval Given Tests (Loop 2)             & 130.40 $\pm$ 2.91                                   & 79.51\%                  \\ 
\#4 & Static Analysis (Loop 3)            & 130.50 $\pm$ 3.11                                   & 79.57\%                  \\  \hline
\multicolumn{4}{c}{Loops 4 and 5 repeated for LLM and \evosuite -generated test cases} \\ \hline

\#5 & LLM Tests   (Loops 4 + 5)                & 132.50 $\pm$ 3.14                                   & 80.55\%                  \\

\#6 & \evosuite Tests   (Loops 4 + 5)          & 132.60 $\pm$ 3.10                                   & 80.85\%                  

\\  \bottomrule
\end{tabular}
}
\label{tab:result}
\vspace{5mm}
\end{table}

\medskip
\textbf{How to run the experiments.}
To run the \humaneval evaluation, the following steps should be followed:

\smallskip
\begin{enumerate}
    \item Build the \textsc{Docker} image and tag it as \texttt{framework}.
    \item Run the \texttt{generate.py} script to begin generating solution attempts for the problems in the dataset.
    \item Once the script completes, the solutions will be in the \texttt{results/result.json} file, and the logs will be in the \texttt{framework\_logs} directory.
\end{enumerate}

\smallskip
There are various scripts  to analyze the data. Each of the analysis scripts takes a list of directories or runs as command-line arguments, each containing the framework JSON logs for the problem. The scripts include:

\begin{itemize}
    \item \texttt{aggregate.py}: Calculates the average number of problems passed at each stage, both individually and cumulatively, with standard deviations.
    \item \texttt{stats.py}: Displays this data as a bar graph.
    \item \texttt{pass\_k.py}: Calculates the pass@k values for the baseline and framework and plots them on a line graph.
\end{itemize}

\medskip  
\textbf{Results.}  
Table~\ref{tab:result} shows the average number of problems solved at each stage of the framework across ten runs. The framework solved an average of 132.6 problems, a 9.2\% improvement over the baseline, which solved 117.5 problems. Average pass@1 increased from 71.65\% for the baseline to 80.85\% after completing all loops of \tool.  

\begin{figure}[t]
    \centering
    \includegraphics[width=0.95\linewidth]{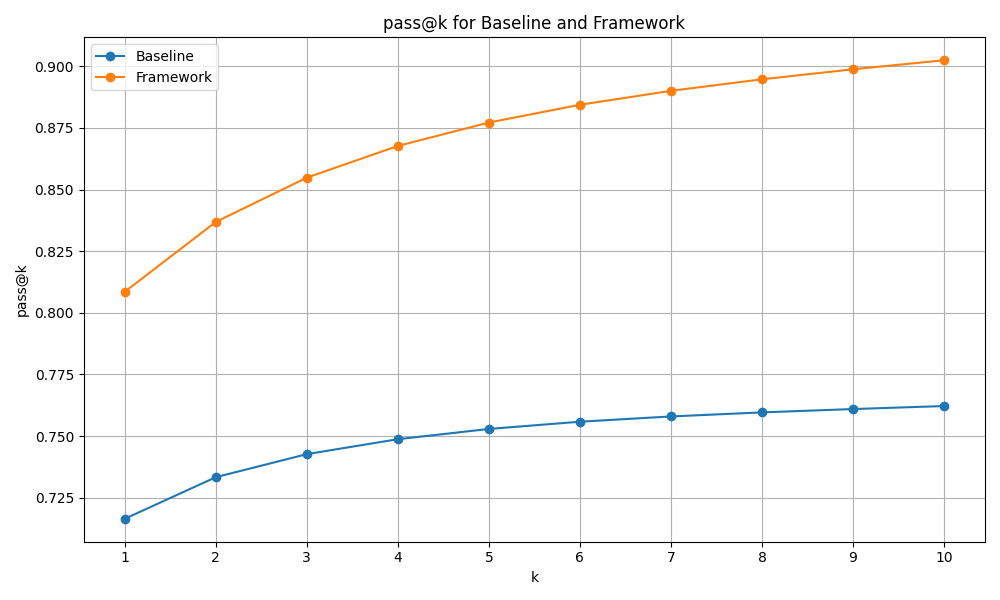}

    \caption{Pass@K for the ten runs, baseline vs \tool}
    \label{fig:passK}
\end{figure}

\smallskip
Figure~\ref{fig:passK} illustrates the improvement in \textit{pass@k} metrics. The baseline achieved a \textit{pass@1} of 71.65\%, rising to 76.22\% at \textit{pass@10}. In comparison, \tool achieved a \textit{pass@1} of 80.85\% and a \textit{pass@10} of 90.24\%.  The framework consistently outperformed the baseline, with a 9.2\% increase at \textit{pass@1} and a peak difference of 14.02\% at \textit{pass@10}. The iterative feedback loops in \tool amplified the baseline’s improvements as the number of code samples increased.  

\smallskip
The results show that the performance of \textsc{GPT-4o-mini} benefited from the iterative refinement provided by \tool. The largest improvement came from the Compilation Loop, which ensures that code compiles successfully, enabling subsequent tests and analyses. The feedback of the \humaneval's example tests (Loop 2) also contributed significantly, aligning the generated code more closely with the problem requirements.  

\smallskip
Later stages of the framework provided smaller gains. The LLM-generated tests (Loop 4) showed (small) improvements, as they could targeted expected behavior~\cite{konstantinou2024llms,Ruberto2025FromImplemented} (rather than implemented one). Differently, \evosuite-generated tests had zero gain in pass@1, as expected. Indeed, \evosuite generates tests with regression oracles to validate implemented logic (for regression testing), rather than checking  program correctness~\cite{konstantinou2024llms,fraser2011evosuite}. 

\section{Limitations and Future Work}

One limitation of \tool is the time required to complete all loops, which increases computational costs due to the frequent invocation of the LLM (often taking several minutes). This makes the approach less \textbf{sustainable}, as repeatedly invoking the LLM is very expensive in terms of both energy and computation. Future work could focus on reducing the number of LLM interactions by prioritizing feedback that is most likely to significantly improve the code. For example, \tool could begin by addressing the compilation error associated with the most issues in the code.

\smallskip
Another important future work is \textbf{evaluating the quality of the tests} generated by \tool (in particular the improvement of Loop 5). Due to space limitations, this was not included in this paper. It will be interesting assessing both LLM-generated and \evosuite tests generated and improved via feedback loops~\cite{ouedraogo2024llms}.

\smallskip

Future work could also integrate \tool as \textbf{a plugin for IDEs} to encourage adoption by developers. While researchers may continue using the command-line version, an IDE plugin would improve usability, bringing \tool closer to a mature, IDE-friendly tool.

\smallskip
Finally, extending \tool to \textbf{support \textsc{Python}} is also important. This could replace \evosuite with \textsc{Pynguin}~\cite{lukasczyk2022pynguin} to generate \textsc{Python} test cases, and PIT with \textsc{MutPy}.

\section{Related Work}

Using feedback loops is a common approach to improving LLM-generated code~\cite{wang2023review}. However, \tool is the first framework designed to reduce the repeated effort typically required to set up such loops. Additionally, \tool introduces new types of feedback loops that are not explored in other techniques, which  focus solely on handling compilation errors and test failures~\cite{schafer2023empirical, liu2024llm, yu2023llm,kang2023large}.

\smallskip
The work most closely related is \textsc{SelfEvolve}~\cite{quoc2024empirical}, which employs a self-correcting loop with a \textsc{Python} interpreter to detect and fix errors in the generated code, including test failures. The loop iterates until all errors are resolved or a maximum number of iterations is reached. However, \textsc{SelfEvolve} is limited to \textsc{Python}  and does not implement feedback loops using static analysis output or automated test generation, as \tool does. Also, \textsc{SelfEvolve} is not currently publicly available, and its companion paper is an ArXiv preprint, which may not have been peer-reviewed yet.

\section{Conclusion}

This paper presented \tool, a framework to automatically improve LLM-generated \textsc{Java} code through iterative feedback loops. By open-sourcing \tool, we aim to empower researchers and developers with an important asset to reduce repeated effort when generating code with LLMs. Moreover, the community could extend its capabilities, optimize its performance, and explore new feedback mechanisms. We hope that \tool will serve as a useful starting point for improving the quality and reliability of LLM-generated code. 

\bibliographystyle{IEEEtran}  
\bibliography{bib}  
\end{document}